\documentclass[preprint2]{aastex}

\slugcomment{To appear in the ApJ Letters}
\received{5/31/2001}
\revised{7/12/2001}
\accepted{7/17/2001}

\makeatletter
\def\ale{\mathrel{\mathpalette\gl@align<}}
\def\age{\mathrel{\mathpalette\gl@align>}}
\def\gl@align#1#2{\lower.6ex\vbox{\baselineskip\z@skip\lineskip\z@
\ialign{$\m@th#1\hfil##\hfil$\crcr#2\crcr\sim\crcr}}}
\makeatletter
\newcommand{\um}{\mu m}

\shorttitle{Deep Near-IR Images of AFGL 618} 
\shortauthors{Ueta, Fong, \& Meixner}

\begin{document}
\title{Westbrook's Molecular Gun: \\
Discovery of Near-IR Micro-Structures in AFGL 618%
\footnote{Based partially on data obtained at the Subaru Telescope, 
which is operated by the National Astronomical Observatory of Japan.}}

\author{Toshiya Ueta, David Fong, and Margaret Meixner}
\affil{Department of Astronomy, MC-221, 
University of Illinois at Urbana-Champaign, 
Urbana, IL  61801\\
ueta@astro.uiuc.edu, d-fong@astro.uiuc.edu, meixner@astro.uiuc.edu}

\begin{abstract}
We present high-sensitivity near-IR images of a carbon-rich 
proto-planetary nebula, AFGL 618, obtained with the
Subaru Telescope.
These images have revealed ``bullets'' and ``horns'' 
extending farther out from the edges of the previously known 
bipolar lobes.
The spatial coincidence between these near-IR micro-structures 
and the optical collimated outflow structure, together with the 
detection of shock-excited, forbidden IR lines of atomic species,
strongly suggests that these bullets and horns represent 
the locations from which [\ion{Fe}{2}] IR lines arise.
We have also discovered CO clumps moving at $> 200$ km s$^{-1}$ 
at the positions of the near-IR bullets by re-analyzing 
the existing $^{12}$CO $J=1-0$ interferometry data.
These findings indicate that the near-IR micro-structures 
represent the positions of shocked surfaces at which fast-moving 
molecular clumps interface with the ambient circumstellar 
shell.
\end{abstract}


\section{Introduction}

AFGL 618 (CRL 618, {\sl IRAS\/} 04395+3601) is a carbon-rich,
extremely evolved proto-planetary nebula (PPN) 
transforming itself into a planetary nebula (PN).
Since its identification \citep{westbrook75},
AFGL 618 has been extensively studied as one of the prime 
examples of a bipolar PPN (e.g., \citealt{cs82,latter92}).
From early observations, the emission from the 
bipolar lobes has been known to have two components:
scattered star light through the biconical openings of
an optically thick dust torus around the central star and 
shock-excited line emission arising from the lobes themselves 
\citep{sc81,kelly92}.

Recent Hubble Space Telescope ({\sl HST\/}) observations
in shock-excited line emission have revealed collimated 
outflow structure in the bipolar lobes \citep{trammell99}.
This evident connection between the bipolar morphology and 
the collimated outflows suggests the presence of very effective 
collimation mechanism(s) at work in the innermost region 
of AFGL 618, possibly deep within the dust torus.
Here we present results from 
high-resolution, high-sensitivity near-IR imaging of 
AFGL 618 using the 8.2 m Subaru Telescope and discuss their 
implications.

\section{Observations}

We observed AFGL 618 using the Infrared Camera and Spectrograph 
(IRCS; \citealt{kobayashi00}) at the Subaru Telescope on 2001 
February 5 (Program ID: S00-072).
The IRCS has two Raytheon ALADDIN II InSb arrays, which give 
$60\arcsec \times 60\arcsec$ field of view at $0\arcsec.058$ 
pix$^{-1}$ without adaptive optics.
The observations were made under marginally clear but very 
gusty conditions.
Seeing varied between $0\arcsec.6$ and $0\arcsec.9$ throughout 
the night.
We employed an on-chip, 5-point dithered pattern to image
and used long exposure times to detect faint nebulosities.
Table 1 summarizes the observations.

Standard IR data reduction and calibration procedures were 
followed, except that the images were extinction-corrected 
using the averaged correction factors for Mauna Kea \citep{kriscinas87} 
because the correction factors were not reliably derived
from the data.
In addition, continuum emission in the H$_{2}$ and Br $\gamma$ 
images was subtracted by estimating continuum emission at these 
bands from the K-continuum image.
Absolute flux calibration errors are estimated to be approximately 
$\pm 10\%$, which reflects the less-than-average sky conditions.
While our {\sl K'\/} magnitude is comparable to the {\sl K} magnitude 
of \citet{westbrook75} despite the recent reports on the increase 
of the integrated flux at {\sl K} \citep{latter92,latter95},
our {\sl J} and {\sl H} magnitudes are brighter than 
previously measured.
The change in the brightness could be intrinsic to the
rapidly evolving nature of the object, however, the quality of 
our data prevents us from elaborating further.

Selected near-IR images of AFGL 618 are presented in Fig. 1.
In these images, any small variations of the surface brightness 
were obscured by very strong emission from the central peak.
Thus, we applied unsharp masks to the original images to enhance
small variations in surface brightness.
Such structure-enhanced images, presented alongside the original 
images, have successfully revealed the clumpy structure of the nebula.

\section{Results and Discussions}

\subsection{Deep Near-IR Images}

The near-IR emission structure of AFGL 618 consists of a highly 
centralized peak and a faint nebulosity that is elongated in 
the E-W direction.
The {\sl J\/} band image clearly shows the bipolarity of the 
source as seen in the optical (e.g., \citealt{westbrook75}).
A similar E-W elongation is seen at {\sl H}\/ (not shown). 
The {\sl K'\/} band image is relatively wider in the N-S
direction compared with the images at {\sl J\/} and {\sl H\/}.
A similar morphological trend has been seen previously and model
calculations have suggested that dust scattering would be the
primary source of the morphological change in the near-IR
\citep{latter92}.
The overall emission structure of the nebula is such that the 
eastern lobe is significantly brighter than the western lobe 
due to the central peak located in the eastern lobe.
This corroborates our current understanding of the structure of 
AFGL 618, in which a bipolar nebula with a wide opening 
angle is inclined at $\sim 45^{\circ}$ with respect to the line 
of sight with its eastern lobe pointing towards us 
(e.g., \citealt{cs82}).

In addition to the previously known bipolar lobes, 
we have detected three ``bullets'' that are isolated 
from the western lobe by about 2{\arcsec} and two 
``horns'' that are connected to the eastern lobe.
Our images present the first clear detection of these 
micro-structures owing to better resolution and
sensitivities.
Previous near-IR images \citep{latter92} seem to show 
some evidence for the eastern horns in the noisy lowest 
contours, but did not cover the region of the sky in 
which the western bullets are present.
The most recent images by \citet{latter95}, while failing 
to show the eastern horns due to the coarse pixel scale 
(0\arcsec.6 pix$^{-1}$), appear to indicate that there 
is a region of extended emission spatially coincident 
with the brightest western bullet.
Although the extension seems to be isolated from the 
western lobe, it was not recognized as a micro-structure.
In the present image at {\sl J}, the brightness of the 
two brighter bullets are at least comparable to
the peak of the western lobe.
In the most recent {\sl J} image \citep{latter95},
however, the tip of the western extension appeared to 
be dimmer than the western lobe and the middle bullet 
was not even seen.
Thus, these bullets may be varying their brightnesses
and/or positions (see \S 3.3) in the past years.
Therefore, continued monitoring of 
these micro-structures will be worthwhile.

The continuum-subtracted narrowband images also show a 
similar elongation in a somewhat different manner.
In H$_{2}$, the central peak is less prominent and,
while the eastern horns can be slightly visible,
there is no apparent western bullets.
Although the H$_{2}$ emission is known to have both thermally 
excited and UV fluorescent components \citep{latter92}, 
our image alone can not isolate these components.
There appear to be faint protrusions emanating from the 
central peak at the 3 $\sigma_{\rm sky}$ level, along the 
equatorial plane defined by the ``dust waist.''
These equatorial structures may be similar to those 
seen in another well-known PPN, AFGL 2688 (the Egg Nebula, 
\citealt{sahai98}).
On the other hand, the Br $\gamma$ image, which originates 
from the compact \ion{H}{2} region around the central star 
\citep{kb84}, is very compact.

Meanwhile, there is a growing interest on a particular 
type of morphology, so-called ``concentric arcs,'' which 
are typically found around evolved stars 
(e.g., \citealt{hrivnak01}).
These arcs appear to be limb-brightened segments of 
spherical shells created by the asymptotic giant branch 
(AGB) wind.
The apparent coexistence of concentric arcs within 
a bipolar nebula as seen, for example, in AFGL 2688 
\citep{sahai98} presents a challenge in formulating a mass 
loss scenario in evolved stars.
Despite the new discoveries of the bullets and horns,
we do not find any apparent evidence for the presence 
of the concentric arcs in our images above one 
$\sigma_{\rm sky}$ level. 

\subsection{Near-IR ``Bullets'' and ``Horns''}

The surface brightness of the bullets and horns is comparable:
the brightest western bullet in {\sl J\/} is actually the second 
brightest point in the entire nebulosity.
This emission structure is unusual because the surface brightness 
due to scattered stellar emission is expected to decrease 
monotonically as the distance from the central star increases.
In such a case, the western (far) side of the nebula 
should appear darker than the eastern (near) side.

In fact, the bullets and horns are spatially coincident 
with the tips of the collimated outflows imaged in shock-excited 
forbidden lines 
([\ion{O}{1}] 6300{\AA} and [\ion{S}{2}] 6717, 6731{\AA})
by {\sl HST} \citep{trammell99}.
The emission clumps seen throughout the nebula in 
the structure-enhanced images agree quite well with the optical 
outflow morphology in the {\sl HST} images.
Moreover, \citet{kelly92} detected [\ion{Fe}{2}] lines at 1.321 
\& 1.328 $\um$ at 4 to 6{\arcsec} east and 4{\arcsec} west 
of the mid-point between the optical bipolar lobes.
These locations respectively coincide with the positions
of the horns and bullets.
\citet{hora99} have, in addition, detected relatively
strong [\ion{Fe}{2}] lines at 1.26 and 1.64 $\um$ from this object.
Since [\ion{Fe}{2}] IR lines are known to be good probes of 
jets and shocks in dense material (e.g., \citealt{gwl87,reipurth00}),
we would conclude that the newly detected near-IR bullets and 
horns are mainly caused by {\sl in situ} shock-excited [\ion{Fe}{2}] 
line emission.

This interpretation would not conflict with the unusual 
emission structure within the nebula because the strength 
of the line emission depends on the shock conditions and 
geometry.
The difference in the micro-structures (bullets vs. horns) 
can be attributed to the different degree of collimation 
of the outflows in the lobes.
The {\sl HST\/} images show a narrower outflow structure in 
the western lobe than in the eastern lobe.
This may imply that the shock surfaces are distributed
over a larger area in the eastern lobe than in the western 
lobe, resulting in equally bright horns and bullets
despite the presumed inclination of the bipolar 
nebula.
While the brightest southernmost bullet is barely visible, 
the middle bullet is the brightest in [\ion{S}{2}] and 
[\ion{O}{1}] \citep{trammell99}.
This can also be attributed to different excitation 
conditions among these bullets.
Future high-resolution spectroscopy would be needed to 
determine shock conditions in these micro-structures.

While the bullets seem to be a unique feature among PPNs,
the horns resemble the ones seen in AFGL 2668 
\citet{latter95,sahai98}.
However, the origins of the horns do not seem to be the same.
In the case of AFGL 618, the horns seem to be a result 
of outflows towards the polar region of the nebula,
whereas the horns in AFGL 2668 is a result of scattering 
\citep{latter95,sahai98}.
This indicates that different mechanisms could form
similar morphologies and thus the nebula shaping 
mechanism may not be uniquely identified from the
morphology alone.

\subsection{Near-IR Bullets vs. Molecular Bullets}

\citet{trammell99} argued, based on the excitation gradient 
in the outflows, that the collimated morphology is a result 
of material flows impinging on the surrounding medium.
In fact, AFGL 618 is known to host one of the fastest molecular 
outflows ever found in the Galaxy (e.g., \citealt{burton86}).
Especially of interest are the single dish observations 
of CO \citep{cernicharo89,gammie89}, in which
the authors have detected outflows at $> 200$ km s$^{-1}$.
The follow-up interferometric observations \citep{neri92,hajian96}, 
however, have detected only medium velocity ($\sim 70$ km s$^{-1}$) 
outflows in the central region.
While \citet{kastner01} have demonstrated that the high 
velocity ($\sim 120$ km s$^{-1}$) H$_{2}$ tends to be found 
closest to the central star,
their position-velocity map also shows the presence of fast 
moving clumps near the edge of the bipolar nebulosity.
In any case, CO outflows at $> 200$ km s$^{-1}$ have not yet 
been spatially identified.

As we have seen, our near-IR images have revealed locations
from which shock-excited [\ion{Fe}{2}] line emission seems 
to arise, and therefore, these bullets and horns may well 
represent the shocked surfaces of high-velocity ``CO bullets'' 
which are penetrating into an ambient AGB shell.
To obtain more kinematical insights, we have re-analyzed 
millimeter interferometry data of the $^{12}$CO $J = 1 - 0$ 
line emission originally presented elsewhere \citep{meixner98}.
In this analysis, we have used only the high-resolution data 
(UV range of $7-91 {\rm k}\lambda$) to spatially filter out 
the extended emission.
Robust weighting of the visibility data yielded a 
$2\arcsec.0 \times 1\arcsec.7$ CLEAN beam with a position angle
of $54.7^{\circ}$.  
This resolution is appropriate for direct comparisons with 
the near-IR images.
Fig. 2 shows the false-color {\sl J\/} band image overlaid 
with CO contours integrated between $-140$ km s$^{-1}$
and 100 km s$^{-1}$.
Also displayed are CO spectra obtained from a 
$2{\arcsec} \times 2{\arcsec}$ patch of the sky at eight 
locations in the nebula.
In these spectra, typical rms noise per 8 km s$^{-1}$ 
channel is 0.07 Jy beam$^{-1}$.

The emission peak spectrum (panel D) shows both blue- and 
red-shifted wings on both sides of the broad peak at 
$\sim 21.5$ km s$^{-1}$ representing the central velocity
of the object \citep{neri92,hajian96}.
A deep dip in the blue-shifted wing is due to self-absorption 
\citep{neri92,hajian96,meixner98}.
The spectra of the lobes (C and E) respectively exhibit 
large blue- and red-shifted wings as expected from the
presumed bipolar outflow.
On the other hand,
the spectra of the horns (A and B) seem to indicate the 
presence of faint broad wings on both sides of the peak 
at the central velocity.
This large velocity dispersion is consistent structure-wise 
with the less-collimated eastern horns as suggested 
by the near-IR emission strengths and the optical outflow 
morphologies. 
The spectra of the bullets (F, G, and H) do not show any 
apparent red-shifted wings, but high-velocity ``clumps'' 
appear to be present.
At the positions of the two brightest bullets (G and H),
we detect emission from clumps moving at about $-220$ km s$^{-1}$.
Although weak in absolute intensity, emission from these clumps 
registers about 4 $\sigma$ in the position-velocity 
diagrams.
In addition, velocity channel maps show consistent structures
over several velocity frames around $-220$ km s$^{-1}$.
Therefore, we conclude that the detections of these high-velocity
CO clumps are real.

There also appear to be CO clumps moving at different 
velocities.
For example, there are clumps moving at about $-140$ and
$-45$ km s$^{-1}$ respectively at F and H.
The existence of an ensemble of fast moving clumps
have already been observed in H$_{2}$ \citep{burton86}.
This coincidence of the fast moving CO clumps and 
shock-excited IR and optical emission regions
strongly suggests that shock excitation is induced 
as these molecular bullets impinge on the ambient
medium.

This is the first spatial identification of the fastest 
moving components of CO emission in AFGL 618.
Fast molecular outflows have been well studied in the context 
of Herbig-Haro (HH) objects. 
\citet{herbst96} observed jets in T Tau at [\ion{Fe}{2}] 
1.644 $\um$ and H$_{2}$ 2.122 $\um$ and attributed the spatial 
displacement between the peaks in these bands to the orientation 
of the jet head with respect to observers.
In their scenario, [\ion{Fe}{2}] line emission comes 
from the head of the outflow where dissociative, fast 
shock is induced and H$_{2}$ emission comes from 
the oblique bow-shocked zone that forms the ``wakes''
behind the advancing molecular bullets.
Comparing our {\sl J\/} and H$_{2}$ images (especially 
the structure-enhanced ones), we immediately see that H$_{2}$ 
emission structure outlines the periphery of the
wakes behind the CO bullets (bullets and horns in {\sl J\/})
as seen in the Orion nebula \citep{allen93}.
To secure the detection of high-velocity molecular
clumps, we need to investigate kinematics of molecular 
species at not only the central part but also the edges of 
the bipolar structure.

\section{Summary}

We have discovered near-IR micro-structures (bullets and horns)
extending farther out from the previously known bipolar
structure of an evolved PPN, AFGL 618.
Based on the emission structure of the micro-structures 
and the results from the previous spectroscopy, we have concluded
that the near-IR micro-structures are likely due to 
line contamination by shock-excited IR lines, especially 
by the [\ion{Fe}{2}] lines.
We have also found the fastest moving ($\sim 220$ km s$^{-1}$) 
CO molecular clumps coincident with the positions of the 
near-IR micro-structures.
These findings strongly suggest that shock-excitation
is being induced by fast-moving CO clumps impacting on
the surrounding medium at the positions of the micro-structures 
located at the edges of the bipolar nebula.
The presence of HH type molecular flows is very 
likely, as suggested from the morphological relationship
between shock-excited line emission (represented by 
near-IR bullets and horns) and H$_{2}$ emission.
It will be worthwhile to continue observing AFGL 618
for its full spatial extent at the highest resolutions 
in the optical and near-IR to see if there really exist
brightness and/or spatial variations
and in molecular lines to better characterize the velocity 
structures and the shock geometry in evolved stars.
Proper understanding of these molecular outflows 
and the resulting shocks is of critical importance for
constructing the theory of structure formation at the 
earliest PN phase.

\acknowledgments
Ueta would like to thank the IRCS support scientist, 
Dr. Hiroshi Terada, and the staff at the Subaru Telescope for 
their support.
An anonymous referee is thanked for valuable comments.
We also acknowledge support from the NSF CAREER award 
AST 97-33697, the Laboratory for Astronomical Imaging at 
the University of Illinois, and NSF grant AST 99-81363.

\begin{deluxetable}{crlcrrrl}
\tablecolumns{7} 
\tablewidth{0pc} 
\tablecaption{Summary of Subaru/IRCS Observations of AFGL 618}
\tablehead{
\multicolumn{3}{c}{Filter} &
\colhead{Exposure} &
\colhead{Size} &
\colhead{Flux} &
\multicolumn{2}{c}{Peak ($1 \sigma_{\rm sky}$)} \\
\colhead{Band} &
\multicolumn{2}{c}{$\lambda$ ($\delta \lambda$) [$\um$]} &
\colhead{(sec)} &
\colhead{($\arcsec \times \arcsec$)} &
\colhead{(mJy)} &
\multicolumn{2}{c}{(mJy arcsec$^{-2}$)}}
\startdata
J &
1.25\phn &
(0.16) &
600, 1000 &
$17.0 \times 6.7$ & 
13 & 
3 & 
(0.003)\\

H &
1.63\phn &
(0.30) &
160 &
$15.8 \times 6.7$ & 
75 & 
44 & 
(0.02) \\

K' &
2.12\phn &
(0.35) &
80 &
$14.2 \times 7.9$ & 
118 & 
63 & 
(0.03) \\

H$_{2}$ &
2.122 &
(0.032) &
300 &
$12.4 \times 5.2$ & 
187 & 
49 & 
(0.2) \\

Br $\gamma$ &
2.166 &
(0.032) &
150 &
$8.7 \times 5.5$ & 
49 & 
11 & 
(0.1) \\

K$_{\rm cont}$ &
2.280 &
(0.032) &
200 &
$8.5 \times 5.5$ & 
193 & 
110 & 
(0.1) \\
\enddata
\end{deluxetable}

\begin{figure}[h]
\plotone{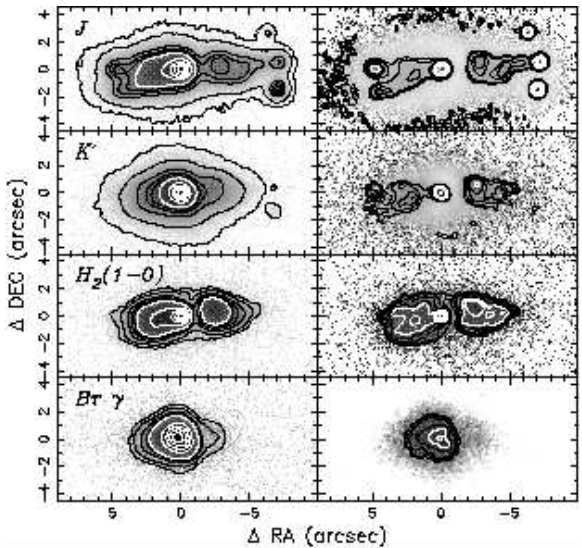} 
\figcaption{Selected near-IR images of AFGL 618 (left)
and their structure-enhanced counterparts (right).
The RA and DEC offsets are shown with respect to the peak.
White contours run from 90\% to 10\% of the peak intensity 
with a 20\% interval, then black 
contours are shown with an arbitrary spacing 
to outline the emission structure.
The lowest contour represents $3 \sigma_{\rm sky}$.
Contours seen well away from the bipolar lobes in the
structure-enhanced images are artifacts in the image 
processing.}
\end{figure}

\begin{figure}[h]
\plotone{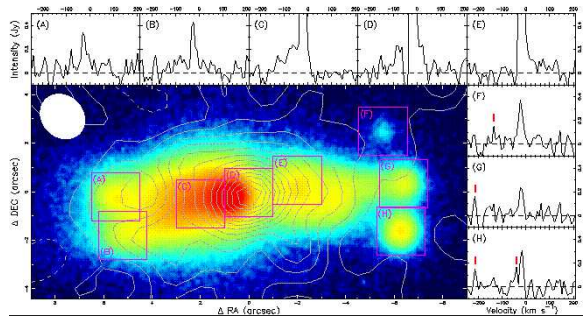}
\figcaption{The false-color {\sl J\/} image overlaid with 
the CO integrated map using the same convention as Fig. 1.
The white ellipse at the top left indicates the beam shape.
CO spectra at various positions are also displayed 
with negative values corresponding to blue-shifted velocities.}
\end{figure}
\end{document}